\documentclass[11pt,twoside]{article}

\usepackage{asp2014}

\aspSuppressVolSlug
\resetcounters

\begin{document}
%\ssindex{protocols!TAP}
%\ssindex{protocols!GKE}
%\ssindex{protocols!GCP}
%\ssindex{observatories!ground-based!Rubin}

\title{Rubin Science Platform on Google: the story so far.}

% Note the position of the comma between the author name and the
% affiliation number.
% Authors surnames should come after first names or initials, eg John Smith, or J. Smith.
% Author names should be separated by commas.
% The final author should be preceded by "and".
% Affiliations should not be repeated across multiple \affil commands. If several
% authors share an affiliation this should be in a single \affil which can then
% be referenced for several author names. If only one affiliation, no footnotes are needed.
% See ManuscriptInstructions.pdf and ASP's manual2010.pdf 3.1.4 for more details

%% Regenerate using:
%%    python $LSST_TEXMF_DIR/bin/db2authors.py > authors.tex

\author{William~O'Mullane,$^1$ Frossie~Economou,$^1$ Flora~Huang,$^2$ Dan~Speck,$^3$ Hsin-Fang~Chiang,$^1$ Melissa~L.~Graham,$^4$ Russ~Allbery,$^1$ Christine~Banek,$^1$ Jonathan~Sick,$^1$ Adam~J.~Thornton,$^1$ Jess~Masciarelli,$^2$ Kian-Tat~Lim,$^5$ Fritz~Mueller,$^5$ Sergey~Padolski,$^6$ Tim~Jenness,$^1$ K.~Simon~Krughoff,$^1$ Michelle~Gower,$^7$ Leanne~P.~Guy,$^1$ and Gregory~P.~Dubois-Felsmann$^8$ }
\affil{$^1$Vera C. Rubin Observatory/NOIRLab, 950 N.\ Cherry Ave., Tucson, AZ  85719}
\affil{$^2$Google Inc., 1600 Amphitheatre Parkway, Mountain View, CA 94043}
\affil{$^3$Burwood Group, 125 South Wacker Drive Suite 2950, Chicago, IL 60606}
\affil{$^4$University of Washington, Dept.\ of Astronomy, Box 351580, Seattle, WA 98195}
\affil{$^5$SLAC National Accelerator Laboratory,  2575 Sand Hill Rd., Menlo Park, CA 94025}
\affil{$^6$Brookhaven National Laboratory, Upton, NY 11973}
\affil{$^7$NCSA, University of Illinois at Urbana-Champaign, 1205 W.\ Clark St., Urbana, IL 61801}
\affil{$^8$IPAC, California Institute of Technology, MS 100-22, Pasadena, CA 91125}
\paperauthor{William~O'Mullane}{womullan@lsst.org}{0000-0003-4141-6195}{Rubin Observatory Project Office}{}{ Tucson}{AZ}{85719}{ USA}
\paperauthor{Frossie~Economou}{}{0000-0002-8333-7615}{Rubin Observatory Project Office}{}{ Tucson}{AZ}{85719}{ USA}
\paperauthor{Flora~Huang}{huangflora@google.com}{None}{Rubin Observatory Project Office}{}{ Tucson}{AZ}{85719}{ USA}
\paperauthor{Dan~Speck}{dspeck@burwood.com}{None}{Rubin Observatory Project Office}{}{ Tucson}{AZ}{85719}{ USA}
\paperauthor{Hsin-Fang~Chiang}{}{0000-0002-1181-1621}{Rubin Observatory Project Office}{}{ Tucson}{AZ}{85719}{ USA}
\paperauthor{Melissa~L.~Graham}{}{0000-0002-9154-3136}{Rubin Observatory Project Office}{}{ Tucson}{AZ}{85719}{ USA}
\paperauthor{Russ~Allbery}{}{None}{Rubin Observatory Project Office}{}{ Tucson}{AZ}{85719}{ USA}
\paperauthor{Christine~Banek}{}{None}{Rubin Observatory Project Office}{}{ Tucson}{AZ}{85719}{ USA}
\paperauthor{Jonathan~Sick}{}{0000-0003-3001-676X}{Rubin Observatory Project Office}{}{ Tucson}{AZ}{85719}{ USA}
\paperauthor{Adam~J.~Thornton}{}{0000-0001-9342-6032}{Rubin Observatory Project Office}{}{ Tucson}{AZ}{85719}{ USA}
\paperauthor{Jess~Masciarelli}{}{None}{Rubin Observatory Project Office}{}{ Tucson}{AZ}{85719}{ USA}
\paperauthor{Kian-Tat~Lim}{}{0000-0002-6338-6516}{Rubin Observatory Project Office}{}{ Tucson}{AZ}{85719}{ USA}
\paperauthor{Fritz~Mueller}{}{None}{Rubin Observatory Project Office}{}{ Tucson}{AZ}{85719}{ USA}
\paperauthor{Sergey~Padolski}{}{None}{Rubin Observatory Project Office}{}{ Tucson}{AZ}{85719}{ USA}
\paperauthor{Tim~Jenness}{}{0000-0001-5982-167X}{Rubin Observatory Project Office}{}{ Tucson}{AZ}{85719}{ USA}
\paperauthor{K.~Simon~Krughoff}{}{0000-0002-4410-7868}{Rubin Observatory Project Office}{}{ Tucson}{AZ}{85719}{ USA}
\paperauthor{Michelle~Gower}{}{None}{Rubin Observatory Project Office}{}{ Tucson}{AZ}{85719}{ USA}
\paperauthor{Leanne~P.~Guy}{}{0000-0003-0800-8755}{Rubin Observatory Project Office}{}{ Tucson}{AZ}{85719}{ USA}
\paperauthor{Gregory~P.~Dubois-Felsmann}{}{0000-0003-1598-6979}{Rubin Observatory Project Office}{}{ Tucson}{AZ}{85719}{ USA}
% Yes they said to have these index commands commented out.
%\aindex{O'Mullane,William}
%\aindex{Economou,Frossie}
%\aindex{Huang,Flora}
%\aindex{Speck,Dan}
%\aindex{Chiang,Hsin-Fang}
%\aindex{Graham,Melissa~L.}
%\aindex{Allbery,Russ}
%\aindex{Banek,Christine}
%\aindex{Sick,Jonathan}
%\aindex{Thornton,Adam~J.}
%\aindex{Masciarelli,Jess}
%\aindex{Lim,Kian-Tat}
%\aindex{Mueller,Fritz}
%\aindex{Padolski,Sergey}
%\aindex{Jenness,Tim}
%\aindex{Krughoff,K.~Simon}
%\aindex{Gower,Michelle}
%\aindex{Guy,Leanne~P.}
%\aindex{Dubois-Felsmann,Gregory~P.}

\begin{abstract}

We describe Rubin Observatory's experience with offering a data access facility (and associated services including our Science Platform) deployed on Google cloud infrastructure as part of our pre-Operations Data Preview program.
\end{abstract}

%Slides on https://docs.google.com/presentation/d/15XiQawtQ4aZCwU1qQ6seXXg6jTfLU7j7K9pjh7ReucE/edit#slide=id.ged99d87c95_0_165
\section{Introduction}

The Legacy Survey of Space and Time \citep{2019ApJ...873..111I} is "deep fast wide" optical/near-IR survey of half the sky in ugrizy bands to r 27.5 (36 nJy) based on 825 visits over a 10-year period.
Carried out by Rubin Observatory on Cerro Pach\'{o}n Chile, the survey will produce around 100\,PB of data consisting of about a billion 16\,Mpix images, enabling measurements for 40 billion objects!

As a big data project, Rubin is providing a "bring-your-compute-to-the-data" model relying on a project-provided Rubin Science Platform (hereafter RSP) to allow science users to access, visualize and analyze survey data products.
This paper describes our experience with an early release of the RSP to users through our pre-operations program.

\section{The Rubin Interim Data Facility on Google cloud}

We decided to deploy services involved in our pre-operations program on a commodity cloud services provider as an Interim Data Facility (IDF) to mitigate uncertainty in the final location of our operations-era US Data Facility (USDF).
Following a competitive tender, we selected Google cloud for this three-year activity to prepare our team for operational readiness and our community for interacting with Rubin data and services.
The IDF is a full stand-alone facility (without dependence on on-premises computing) and is a real high-availability production environment for all services currently oriented to external users. These include the Rubin Science Platform as well as underlying services such as Qserv (our high performance database system) and Butler (data abstraction middleware).

The architecture is depicted in Figure \ref{fig:IDFarch} and leverages the Google Kubernetes Engine (GKE) and Google Cloud Storage (GCS) to scale up and down as needed depending on demand.
Kubernetes is our reference deployment platform both on-premises and in the cloud as our services make heavy use of container orchestration patterns and consume containers produced by our build systems based on Jenkins and GitHub Actions \citep[see e.g.,][]{2018SPIE10707E..09J}.

Management of Google cloud resources is fully captured in code (Terraform) developed by the Burwood Group and no production services rely on ad-hoc interactive use of the Google console. We use continuous deployment techniques to manage our services including tools such as ArgoCD\footnote{\url{https://argoproj.github.io/cd/}} even though we don't practice pure continuous deployment, having instead settled on a gated release model \citep{SQR-056}.

\articlefigure[width=0.9\textwidth]{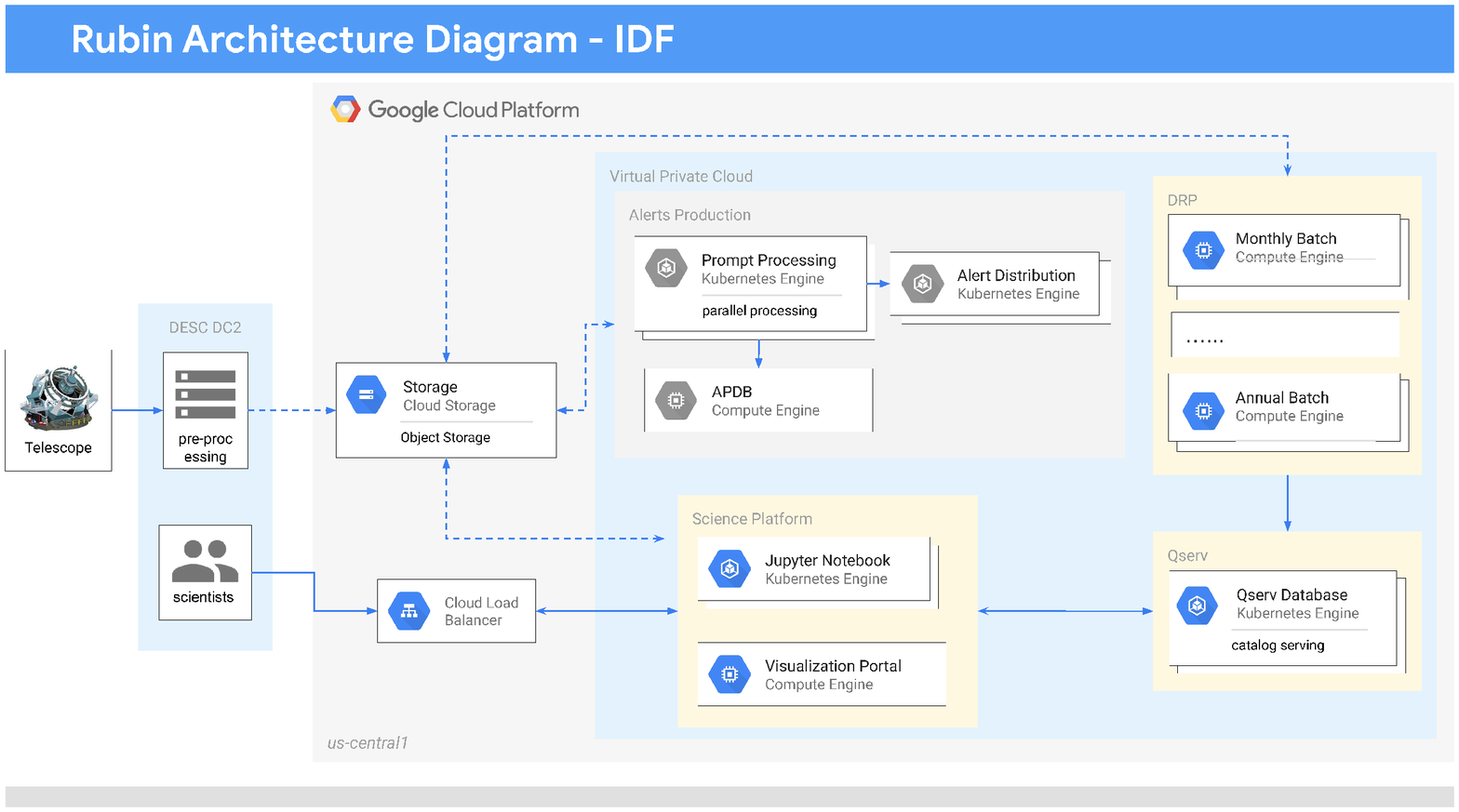}{fig:IDFarch}{Architecture of the Interim Data Facility on Google, gray Items (alerts) are not yet implemented but shall be in 2022.}

\section {The Pre-operations Data Preview programme}
The IDF is hosting what is the first of a series of Data Previews (DP). DP0 is using simulated data (as the telescope has not yet achieved First Light), DP1 will offer commissioning camera data,
and DP2 LSSTCam commissioning data.
% this one ref makes it too long \citep{RDO-011}.
DP0 currently has two phases~\citep{RTN-001} with one focusing on data access and the other on data processing readiness.
In June 2021 \href{https://data.lsst.cloud}{data.lsst.cloud} was opened up to the first 200 delegates selected as part of Data Preview 0.1 with data from the DC2 simulated dataset provided by the DESC collaboration \citep{arXiv:2010.05926}.
For Data Preview 0.2 we shall reprocess the DESC data published "as is" in DP0.1 with the LSST Science Pipelines v23, to generate a fully self-consistent data release that matches the LSST data model.
This will entail producing many types of the final data products and catalogs as well as loading the  generated catalogs in Qserv and will demonstrate readiness (and on-prem/cloud portability) for a number of systems including our PanDA-based processing system and the Generation 3 version of the Butler \citep{2019ASPC..523..653J}.

\section{Community Engagement }

Besides the technical and operational goals, a key aim of deployment of the Interim Data Facility was the engagement of the future community of scientists and students by providing early access to ``DP0 delegates'',a group of users representing the broad science community as learners, testers, and providers of feedback, and tasked with sharing the benefits of their DP0 participation with their communities.

In addition to extensive documentation at \href{https://dp0-1.lsst.io}{dp0-1.lsst.io} supporting delegates in a number of ways, delegate resources included biweekly virtual ``Delegate Assemblies'' with hands-on demonstrations and breakout rooms to facilitate co-working; a dedicated category in the Rubin Community Forum (\url{https://community.lsst.org}, based on the Discourse web forum platform) where delegates can ask questions and discuss their DP0 work; and a GitHub repository where delegates can contribute to shared code and notebooks.
GitHub Issues were also used as the primary means for delegates to request technical assistance. All of the delegate pedagogical resources are publicly accessible.

Some aspects of the Community Engagement model as applied to DP0 are designed to scale to thousands of users for the future Data Previews, such as the extensive documentation, the tutorials, and the use of the Community Forum.
Other aspects, such as the frequent live virtual sessions, may not scale well to thousands of users and were specifically designed to ramp up skill levels from novice to intermediate in order to seed expertise with Rubin software in the community.
For future Data Previews, the Community Engagement strategy will focus on building infrastructure to foster a vibrant science community, that enables self-help, peer-to-peer help and crowdsourced support.

\section{Cloud as a primary platform for astronomical data services}

The successful launch of our Data Preview program with our cloud-based Science Platform reinforced our view that commodity infrastructure offers unparalleled technical and programmatic advantages in delivering large-scale data services to astronomers. Some key points:

\begin{itemize}

    \item Architecting systems using cloud-native techniques (such as targeting Kubernetes as the deployment platform) allows painless transition between commodity and on-premises infrastructure.
    \item There are tangible benefits to working with a highly popular toolchain both for a project and its developers, including the ease of contracting short-term bootstrapping help as Rubin did with the Burwood Group.
    \item Developers love the self-serve aspect of cloud infrastructures (and their velocity reflects this).
    \item Use of commercial cloud environments offers an ideal way to mitigate schedule and technical risk in the delivery of on-premises computing.
    \item There are significant security advantages in placing public users in a separate security domain from on-premises resources.
    \item We are seriously evaluating whether an on-prem/cloud hybrid model is actually the best way forward permanently for Rubin Operations.
\end{itemize}

While use of these services are not cheap (despite academic and government discounts that can be had for the asking), they offer great value for money (and are frequently disadvantaged in comparisons with on-premises cost by the lack of total cost accounting of in-house facilities).
Management buy-in is a big part of any cloud adventure; in our case we were helped by our proof of concept studies with both Amazon \citep{2020arXiv201106044B,DMTN-137} and Google \citep{DMTN-125}. Each ran for six months and demonstrated we could deploy and operate on either platform. This allayed concerns about vendor lock-in and improved our cost models.

We believe use of commodity cloud infrastructure will allow scientific facilities to focus on fertile domain-relevant middleware development instead of on-premises compute facilities and we hope that the funding landscape can evolve to facilitate this paradigm shift.
\bibliography{O3-2}  % For BibTex
\noindent {\tiny This material or work is supported in part by the National Science Foundation through Cooperative Agreement AST-1258333 and Cooperative Support Agreement AST1836783 managed by the Association of Universities for Research in Astronomy (AURA), and the Department of Energy under Contract No. DE-AC02-76SF00515 with the SLAC National Accelerator Laboratory managed by Stanford University.
}
\end{document}